# ZrTe$_2$/CrTe$_2$: an epitaxial van der Waals platform for spintronics

Yongxi Ou[1], Wilson Yanez[1], Run Xiao[1], Max Stanley[1], Supriya Ghosh[2], Boyang Zheng[1], Wei Jiang,[3] Yu-Sheng Huang[1], Timothy Pillsbury[1], Anthony Richardella[1], Chaoxing Liu[1], Tony Low[3,4], Vincent H. Crespi[1], K. Andre Mkhoyan[2], Nitin Samarth[1*]

[1]Department of Physics, The Pennsylvania State University,

University Park, Pennsylvania 16802, USA

[2]Department of Chemical Engineering and Materials Science,

University of Minnesota, Minneapolis, Minnesota 55455, USA

[3]Department of Electrical & Computer Engineering,

University of Minnesota, Minneapolis, Minnesota 55455, USA

[4]School of Physics & Astronomy,

University of Minnesota, Minneapolis, Minnesota 55455, USA

*Corresponding author. Email: nsamarth@psu.edu

**The rapid discovery of two-dimensional (2D) van der Waals (vdW) quantum materials has led to heterostructures that integrate diverse quantum functionalities such as topological phases, magnetism, and superconductivity. In this context, the epitaxial synthesis of vdW heterostructures with well-controlled interfaces is an attractive route towards wafer-scale platforms for systematically exploring fundamental properties and fashioning proof-of-concept devices. Here, we use molecular beam epitaxy to synthesize a vdW heterostructure that interfaces two material systems of contemporary interest: a 2D ferromagnet (1T-CrTe$_2$)**

**and a topological semimetal (ZrTe₂). We find that one unit-cell (u.c.) thick 1T-CrTe₂ grown epitaxially on ZrTe₂ is a 2D ferromagnet with a clear anomalous Hall effect. In thicker samples (12 u.c. thick CrTe₂), the anomalous Hall effect has characteristics that may arise from real-space Berry curvature. Finally, in ultrathin CrTe₂ (3 u.c. thickness), we demonstrate current-driven magnetization switching in a full vdW topological semimetal/2D ferromagnet heterostructure device.**

**Main**

Van der Waals (vdW) materials are an exciting playground for the discovery of emergent behavior in electrical, optical, and thermal properties in the two-dimensional (2D) limit and are potentially attractive for next-generation device applications [1–8]. The recent demonstration of long-range ferromagnetic order in 2D vdW materials has opened another new avenue to study magnetism in atomically thin films [9–15]. While many studies of 2D vdW ferromagnets have focused on mechanically exfoliated flakes [16–19], 2D vdW ferromagnets embedded in heterostructures create new opportunities for manipulating and engineering magnetic properties [20–22]. Such multilayer structures can potentially serve as building blocks for 2D magnetic and spintronics applications [14,15]. For example, chiral magnetic textures have been observed in mechanically stacked heterostructures using the vdW ferromagnet Fe₃GeTe₂ [23,24]. Spin-orbit torque (SOT)-assisted magnetization switching has also been reported in layered vdW ferromagnets interfaced with heavy metals [25–27].

Amongst the candidate 2D ferromagnets, 1T-CrTe₂ has an interesting combination of properties. Bulk 1T-CrTe₂ is a known ferromagnetic material with a Curie temperature, $T_c$, above room temperature. This persists even in flakes exfoliated down to thicknesses of tens of nanometers [28–31]. Synthesized thin films of this material also show a relatively high $T_c$ down to the quasi-2D regime [32,33]. An in-plane-to-out-of-plane transition of the magnetic easy axis in 1T-CrTe$_2$ may be controlled through thickness and strain in thin films [34,35]. Finally, 1T-CrTe$_2$ single crystals show reasonable stability against degradation after being exposed to atmosphere [34].

Here, we report the synthesis by molecular beam epitaxy (MBE) of full vdW heterostructures that interface ultrathin ferromagnetic 1T-CrTe$_2$ films with ZrTe$_2$, a candidate topological Dirac semimetal. We use *in-vacuo* angle-resolved photoemission spectroscopy (ARPES) to measure the band dispersion of metallic 1T-CrTe$_2$ and find that it is consistent with first-principles calculations. Measurements of the anomalous Hall effect (AHE) demonstrate robust ferromagnetism in both single 1T-CrTe$_2$ epilayers grown on sapphire and in vdW sapphire/ZrTe$_2$/CrTe$_2$ heterostructures. In thick films (12 u.c.) of 1T-CrTe$_2$ layers grown directly on sapphire or on ZrTe$_2$, we observe an AHE whose magnetic field dependence is suggestive of real-space Berry curvature effects. We further use the AHE to demonstrate the persistence of ferromagnetic order in one unit cell of 1T-CrTe$_2$ grown epitaxially on ZrTe$_2$, thus realizing a wafer scale spintronics platform that epitaxially interfaces a 2D ferromagnet with a topological semimetal. Finally, we demonstrate current-induced magnetization switching in an ultrathin full vdW ZrTe$_2$/CrTe$_2$ heterostructure device, where the spin-orbit torque efficiency of the ZrTe$_2$ layer is evaluated via spin-torque ferromagnetic resonance (ST-FMR) measurements in ZrTe$_2$/Ni$_{80}$Fe$_{20}$ (permalloy, Py) heterostructures.

**MBE growth and characterizations of the ZrTe$_2$/CrTe$_2$ heterostructures**

Single-layer 1T-CrTe$_2$ and ZrTe$_2$/CrTe$_2$ (Fig. 1a) heterostructures were grown on (001) sapphire substrates by co-deposition from Cr (Zr) and Te sources in an MBE chamber with a base pressure of ~1×10$^{-10}$ mbar. Growth was monitored with 13 keV reflection high energy electron diffraction (RHEED). Sharp streaky RHEED patterns (Fig. 1b) indicated the epitaxial growth of the materials (see Materials and Methods for details). The film thickness during growth was controlled by the deposition time (~1 u.c. per 10 minutes for CrTe$_2$ growth, ~1 u.c. per 25 minutes for ZrTe$_2$) as calibrated from x-ray reflectometry. To protect the thin films from oxidation during *ex-situ* characterization, we deposited a capping layer of ~40 nm Te. The 1T-CrTe$_2$ crystal structure belongs to the $P\bar{3}m1$ space group (Fig. 1a). We characterized the crystalline structure of the ZrTe$_2$/CrTe$_2$ heterostructures using aberration-corrected scanning transmission electron microscopy (STEM), as shown in the cross-sectional high-angle annular dark-field (HAADF) images in Fig. 1c, revealing an atomically flat interface between ZrTe$_2$ and CrTe$_2$. The atomic alignment between the CrTe$_2$ layers matches well to that of 1T phase CrTe$_2$ (Fig.1c). We used energy dispersive X-ray (EDX) spectroscopy to determine the relative concentration of Cr and Te in the CrTe$_2$ thin films, where it showed Cr/Te=0.53, indicating limited (if any) Cr intercalation (See Supplementary Materials). The lattice constants of the 1T-CrTe$_2$ thin films grown on sapphire, $a = b = 3.93 \pm 0.06$ Å and $c = 6.02 \pm 0.05$ Å ($<90°\times 90°\times 120°>$), were further evaluated from the atomic HAADF-STEM images using sapphire as a reference. The measured in-plane lattice constant is ~3% larger than reported in bulk crystals [28], possibly because our epitaxial layers are strained. Figure 1d shows the topography of 1 u.c. CrTe$_2$ grown on ZrTe$_2$ measured via *in vacuo* transfer to a scanning tunneling microscope (STM). The height profile indicates the thickness ~6.0 Å for the 1 u.c. CrTe$_2$ sample, in good agreement with the TEM results.

Figure 1e shows the x-ray diffraction spectrum of a CrTe$_2$ thin film with peaks corresponding to the out-of-plane (001) growth direction as well as peaks from the sapphire substrate. The rocking curve (inset of Fig.1e) of the CrTe$_2$ (001) peak gives a FWHM ~ 0.03 degree, indicating a reasonable crystallinity in the grown films. Additional reciprocal lattice maps were used to characterize the mosaic spread (see Supplementary information). We also used X-ray photoemission spectroscopy (XPS) to determine the sample composition. Figure 1f shows an interference between Te 3d and Cr 2p XPS spectrum. The presence of chromium was also confirmed by the weak Cr 3s and Cr 2s peaks (not shown here) as well as curve fitting of reference telluride and Cr° spectra acquired under similar conditions. Peak positions obtained were as follows: Te $3d_{5/2}$ at 572.0 eV, Te $3d_{3/2}$ at 582.4 eV, Cr $2p_{3/2}$ at 573.5 eV and Cr $2p_{1/2}$ at 582.7 eV respectively. The XPS element analysis gives the Te/Cr concentration ratio ~2:1, which is in good agreement with the STEM-EDX analysis.

We measured the band structure of the 1T-CrTe$_2$ thin films through *in vacuo* transfer to an ARPES chamber with excitation from the 21.2 eV I$\alpha$ spectral line of a helium plasma lamp. Photoemitted electrons were detected by a Scienta Omicron DA 30L analyzer with 6 meV energy resolution. Figure 2a shows the hexagonal Brillouin zone of 1T-CrTe$_2$ and Fig. 2b shows the band dispersion of the 1T-CrTe$_2$ (001) surface measured along the $\bar{\Gamma}-\bar{M}$ direction at room temperature. The location of the chemical potential within the valence bands indicates the sample is p-type, a fact also confirmed using Hall effect measurements. We performed first-principles density functional theory (DFT) calculations for bulk 1T-CrTe$_2$ including spin-orbit coupling with the magnetic moment oriented out-of-plane, obtaining the band structure shown in Fig. 2c. There is good agreement between the ARPES spectrum and the calculated bands (see Supplementary information for more 1T-CrTe$_2$ calculations). The asymmetry observed in the ARPES intensity is likely due to matrix element effects [36].

Figures 2d and 2e show the ARPES spectra of certain ZrTe$_2$/CrTe$_2$ heterostructures. The ARPES spectrum of ultrathin 1 u.c. CrTe$_2$ grown on ZrTe$_2$ largely resembles the band dispersion of ZrTe$_2$. We attribute this to the CrTe$_2$ 1 u.c. layer (~0.6 nm) being thinner than the mean free path of photoemission electrons near the sample surface, rendering the ARPES signal from the $ZrTe_2$ layer beneath still detectable. Note the linear dispersion of the Dirac band from $ZrTe_2$, supporting the presence of a topological Dirac semimetal phase of the 4 u.c. $ZrTe_2$ film [37,38]. As the $CrTe_2$ film becomes thicker (3 u.c. $CrTe_2$), its ARPES band dispersion looks more similar to the 12 u.c. $CrTe_2$ results. The smooth transition of the band dispersion in the $ZrTe_2/CrTe_2$ heterostructure suggests an excellent epitaxy between the two materials and lays the foundation for the observed robust ferromagnetic order in such bilayers as we discuss below.

**Anomalous Hall resistance measurements**

Next, we describe transport measurements of 1T-$CrTe_2$ epilayers and $ZrTe_2/CrTe_2$ heterostructures. We measured the Hall resistance of the samples as a function of an out-of-plane magnetic field at various temperatures (details of the longitudinal resistivity measurements are given in the Supplementary Materials). Figure 3 shows the sample schematics as well as the Hall resistance of the single layer and heterostructures (see Methods for measuring and analyzing the Hall resistance data). Starting with the results in a 12 u.c. (~7.2 nm thick) $CrTe_2$ film grown on sapphire, the Hall effect at room temperature shows a small nonlinear Hall signal at low magnetic fields, suggesting weak ferromagnetic order with a Curie temperature ($T_c$) in the vicinity of room temperature. This is consistent with an earlier report that a 10 nm thick exfoliated flake of 1T-$CrTe_2$ has a $T_c$ above room temperature [29]. Note that earlier reports indicate that $T_c$ may be enhanced in thin films of $CrTe_2$ and other CrTe compounds in CVD-grown samples as compared to thicker films [34,39].

As we cool the 12 u.c. $CrTe_2$ thin film below 200 K, a stronger AH resistance appears with a hysteresis loop whose coercivity continually increases as the temperature falls below 100 K. The hysteretic AHE loop indicates that the 1T-CrTe$_2$ thin film exhibits an out-of-plane magnetic easy axis. As we discuss later, we observed an out-of-plane magnetic easy axis in all our CrTe$_2$ thin films down to the 1 u.c. limit. This contrasts with the in-plane easy axis observed in bulk exfoliated CrTe$_2$ flakes [29–31], but agrees with other thin-film results [32,34]. To understand the magnetic anisotropy behavior in our samples, we used DFT to compute the magnetic anisotropy energy of 1T-CrTe$_2$ thin films with different lattice constants (see Supplementary Materials). We find that the easy axis is sensitive to both lattice strain, in agreement with a previous prediction [35], and Fermi-level position, making it possible to have different easy axis directions in different experimental settings.

As the sample temperature further decreases below around 40 K, the hysteresis loop of the AHE is replaced by an unconventional shape that has a non-monotonic dependence on magnetic field, vanishing at both low and high field and with a peak at an intermediate field. This feature becomes more prominent at low temperature (2 K). This unusual feature in the Hall resistance has been reported in other ferromagnetic systems where the ferromagnetic order can be of intrinsic origin [40–42] from ferromagnetic doping [43,44] or from an interfacial proximity effect [45]. It is regarded as a signature of the topological Hall effect (THE), arising from complex real space chiral domain structures such as skyrmions [46,47]. This unconventional Hall effect is also seen in CrTe$_2$ films of the same thickness grown on ZrTe$_2$ (Fig. 3c). We speculate that the unusual Hall effect at low temperature in thick CrTe$_2$ films arises because of the presence of a non-collinear inversion-symmetry-breaking Dzyaloshinskii-Moriya interaction. Exfoliated CrTe$_2$ flakes have been shown to exhibit Néel-type domain walls due to the six-fold crystalline symmetry [31]. Since our CrTe$_2$ thin films exhibit an out-of-plane magnetic easy axis, we posit that these Néel-type domain walls transform into a chiral magnetic texture, producing the observed behavior in the Hall measurement.

At this stage, we cannot definitively rule out alternative scenarios such as competing ferromagnetic phases that produce AHE of opposite sign different sign. However, detailed analysis of the variation of the Hall effect as a function of temperature and magnetic field suggests that such a 'trivial' alternative scenario is unlikely (see Supplementary Materials for more discussion). Direct real-space experimental evidence of the chiral domain structures, such as low temperature magnetic force microscopy or Lorentz transmission electron microscopy, will be needed to definitively determine the nature of the magnetic ordering and its impact on the Hall effect.

We now discuss the ferromagnetism in thinner $CrTe_2$ films, focusing on sapphire/$ZrTe_2$/$CrTe_2$/Te heterostructures that are of higher structural quality than ultrathin $CrTe_2$ films grown directly on sapphire. Our measurements show that 3 u.c. thick $CrTe_2$ films only begin to show an AHE below 150 K (Fig. 3d). Measurements down to 2 K show a conventional AHE whose magnitude increases monotonically with decreasing temperature; we do not observe any unconventional signatures in the AHE unlike in 12 u.c. thick $CrTe_2$ films. We note that a thickness-dependent THE has been observed in other material systems [41,42]. In our $CrTe_2$ films, we tentatively attribute the observed thickness dependence of the Hall effect to the enhancement in magnetic anisotropy which prefers an out-of-plane easy axis that arises from a weakening of the Coulomb screening effect in the 2D limit [34]. We speculate that the 3 u.c. $CrTe_2$ film in this case may have a different anisotropy energy compared to the thicker sample, such that it no longer satisfies the anisotropy requirement to form a proper phase to exhibit chiral magnetic texture.

To test the robustness of the ferromagnetic order in the true 2D limit of 1T-$CrTe_2$, we also measured a heterostructure consisting of only 1 u.c. of $CrTe_2$ grown on $ZrTe_2$ (Fig. 3e). Like the 3 u.c. $CrTe_2$ film, an AHE signal appears at around 150 K. The observation of the AHE unambiguously demonstrates the existence of long-range ferromagnetism even down to the 1 u.c.

limit of CrTe$_2$ in the heterostructure, confirming vdW 1T-CrTe$_2$ as a 2D ferromagnet. In contrast to the thicker films, the magnitude of the AHE shows a non-monotonic dependence on temperature. As the temperature is lowered from 150 K, the magnitude initially increases, reaching a maximum at around 100 K. It then decreases with further lowering of temperature, vanishing at around 40 K and then increasing again but with opposite sign. This sign reversal of the AHE may be a consequence of the variation of the Berry curvature [48] induced by the charge transfer between ZrTe$_2$ and CrTe$_2$. A similar phenomenon has reported previously in magnetic topological insulator heterostructures [49]. Figure 3f summarizes the measured temperature-dependent anomalous Hall conductivity of the various CrTe$_2$ samples.

**Current-induced spin-torque and magnetization switching**

Finally, we examine a proof-of-concept spintronic device demonstration of our wafer-scale vdW Dirac semimetal/2D ferromagnet. By using optical lithography, we fabricated 5 μm × 10 μm scale Hall bar devices of a ZrTe$_2$(8 u.c.)/CrTe$_2$(3 u.c.) heterostructure and used the AHE to probe the response of the CrTe$_2$ magnetization to current flowing in the heterostructure (Fig. 4a). We note that the ZrTe$_2$ and CrTe$_2$ films have similar conductivities so that current flows in parallel in the two layers. Figure 4b show magnetization switching at 50 K under a pulsed longitudinal current with an external field 700 Oe parallel to the current direction. An out-of-plane magnetic field was first applied to align the magnetization of CrTe$_2$ to an initial state before using a pulsed current to switch the magnetization (see Methods). As shown in Fig. 4b, positive and negative current pulses switch the magnetization between two states; the switching edge appears to be step-like, indicating the switching likely involves multi-domain nucleation and expansion under the current-induced spin-orbit torque (SOT) from ZrTe$_2$. The average current density for the onset of the magnetization switching of this ZrTe$_2$(8u.c.)/CrTe$_2$(3u.c.) device is about 1.8x10$^7$ A/cm$^2$, which is comparable to the current density needed to switch the magnetization of 3D ferromagnets (e.g. Co, CoFeB) using heavy metals [50,51]. However, we emphasize that because of the complicated nature of the domain nucleation and domain wall motion during the magnetization switching process in micron-meter size devices, it is difficult to directly evaluate the SOT from the switching current density alone [27,51]. We note that we have observed current-switching SOT switching over the temperature range 10 K < T < 90 K without much variation in the threshold switching current density (see Supplementary Material for results at additional temperatures and bias fields). Technical constraints prevent us from directly measuring the SOT efficiency using techniques such as ST-FMR and spin pumping at the low temperature required by the Curie temperature of the 3 u.c. CrTe$_2$. For a better understanding of the efficiency of ZrTe$_2$ as a SOT material, we instead carried out ST-FMR measurements at room temperature on a ZrTe$_2$/permalloy (Py) heterostructure. Figure 4c shows the ST-FMR data measured at room temperature on a 50 μm x 50 μm ZrTe$_2$(5nm)/Py(4nm) device (see Methods). This mixing voltage ($V_{mix}$) signal is the result of the dynamics of the magnetization of the Py layer driven by the SOT induced in the ZrTe$_2$ layer by the input RF current. The resonance shape of $V_{mix}$ can be separated into a symmetric (S) and antisymmetric (A) Lorentzian component respectively, where the symmetric (antisymmetric) component is proportional to the current-induced SOT (Oersted field). The SOT efficiency, $\xi_{FMR}$, defined as the ratio of the spin current ($J_s$) to the charge current ($J_c$), is evaluated from the ratio of the symmetric and antisymmetric components [52]: $\xi_{FMR} = \frac{2e}{\hbar}\frac{J_s}{J_c} = \frac{S}{A}\frac{e\mu_0 M_s t_{ZrTe_2} t_{Py}}{\hbar}\left[1+\left(\frac{M_{eff}}{H_{Res}}\right)\right]^{1/2}$,

where $e$ is the charge of the electron, $\hbar$ is the reduced Planck constant, $\mu_0$ is the permeability of free space, $M_S$ is the saturation magnetization of Py, $t_{ZrTe2(Py)}$ is the thickness of the ZrTe$_2$(Py) layer, $M_{eff}$ is the effective magnetization and $H_{Res}$ is the resonance field respectively. From the data in Fig. 4c, we obtain $\xi_{FMR} = 0.014 \pm 0.005$ for ZrTe$_2$/Py. With the measured electrical conductivity of ZrTe$_2$, $\sigma_{xx}^{ZrTe_2} = 3.16 \times 10^5 \, \Omega^{-1} m^{-1}$, the effective spin Hall conductivity (SHC) of such a ZrTe$_2$ thin film is estimated as: $\sigma_{SH,effetive}^{ZrTe_2} = (\hbar/2e)\sigma_{xx}^{ZrTe_2}\xi_{FMR} \simeq (\hbar/e) 2.2 \times 10^3 \, Sm^{-1}$. While the ST-FMR result clearly demonstrates charge to spin conversion in the Dirac semimetal ZrTe$_2$ layer and shows that the spin current generated in the ZrTe$_2$ layer is playing an important role in the current induced magnetization switching experiment in ZrTe$_2$/CrTe$_2$ heterostructures, the SOT efficiency deduced in this manner is usually only a lower bound of the full SOT efficiency generated inside the SOT material due to the non-ideal interface and interfacial spin transparency. Since the interface in ZrTe$_2$/CrTe$_2$ is a more coherent one compared to that in ZrTe$_2$/Py, we expect a more efficient spin current transfer in the former because of the epitaxial interface and the smooth transition of the band structure as indicated by our ARPES measurements (Fig. 2).

To obtain further insights into the spin-charge conversion generated by ZrTe$_2$, we also carried out first-principles calculations of the SHC of both bulk and multilayer ZrTe$_2$ (Figs. 4d and 4e). For the bulk phase, the SHC near the Fermi level is almost zero, and increases in magnitude with electron or hole doping, which is consistent with our relatively small effective SHC of ZrTe2 determined via ST-FMR (see above). On the other hand, for the trilayer case, the Fermi level shifts to a higher energy with a noticeable broad SHC peak (Fig.4e). The origin of this SHC peak can be attributed to the presence of topological Dirac nodes in vicinity of the Fermi level. We have also calculated the SHC for monolayer and bilayer ZrTe$_2$. The bilayer shows similar behavior as a trilayer while the monolayer shows quite different behavior (See Supplementary Materials). Note that the actual Fermi level of ZrTe$_2$ may be influenced by the presence of the Py layer, due to charge transfer that results from a difference in their work functions. Nevertheless, due to the broad SHC peak in trilayer ZrTe$_2$, we anticipate the Fermi level to be within this conductivity peak.

**Conclusions**

We used MBE to synthesize 1T-CrTe$_2$ thin films and full vdW wafer scale 1T-CrTe$_2$/ZrTe$_2$ heterostructures. The out-of-plane magnetic easy axis of these MBE grown ferromagnetic CrTe$_2$ films is of particular interest for studying proof-of-concept spintronics applications such as perpendicular magnetic tunnel junctions (*50)* and will be technologically relevant once a robust room temperature ferromagnetic state is realized. We observed behavior consistent with a THE in both CrTe$_2$ epilayers and heterostructures as indicated by AHE measurements, suggesting CrTe$_2$ as a promising material platform for studies of chiral magnetic domain structures. We also demonstrated that long-range ferromagnetic order persists in the heterostructure ZrTe$_2$/CrTe$_2$ down to the 1 u.c. limit of 1T-CrTe$_2$, and we further demonstrated current-induced magnetization switching in an ultrathin ZrTe$_2$/CrTe$_2$ full vdW heterostructure and characterized the SOT from ZrTe$_2$ via ST-FMR measurements. The wafer scale epitaxial synthesis of heterostructures that cleanly interface the vdW 2D ferromagnet CrTe$_2$ with the topological semimetal ZrTe$_2$ may provide new opportunities in studying the coexistence of the 2D ferromagnetic and topological phases, interfacial interactions such as proximity induced magnetism in vdW topological semimetals, as well as the SOC interaction and spin torque phenomena in the true 2D limit.

**Methods**

**Sample growth**

We deposited single layer 1T-CrTe$_2$ thin films and ZrTe$_2$/1T-CrTe$_2$ thin film heterostructures using MBE in a Scienta Omicron EVO50 system under ultrahigh vacuum (~10$^{-10}$ mB). The sapphire (0001) substrates were outgassed at 600 °C *in situ* for 1 hour to clean the surface before the deposition of the thin films. The epitaxial 1T-CrTe$_2$ was grown at a substrate temperature of 280 °C via co-evaporation of Te (purity: 99.99%, Alfa Aesar) and Cr (purity: 99.997%, Alfa Aesar) respectively, with the flux ratio ~ 40:1. ZrTe$_2$ was grown at a substrate temperature of 420 °C. Te was sublimated at a significant overpressure compared to the Zr (purity: grade 702, Kurt J.Lesker) which was evaporated via e-beam at a deposition rate of roughly 0.3 Å/min. During the growth of ZrTe$_2$, the film was annealed periodically throughout the growth under a constant tellurium flux in order to avoid vacancies and mitigate defects. The outgassing and growth temperatures were measured by an infrared camera with emissivity of 0.7. RHEED was monitored using a 13 keV electron gun during the growth of the samples. Before *ex situ* characterization and measurements, we capped the samples with 40 nm Te to avoid degradation.

**STM and ARPES measurements**

*In-situ* topography was measured at 300 K after transferring MBE-grown samples *in vacuo* to a Scienta Omicron LT NANOPROBE STM system. We also carried out ARPES measurements at 300 K after *in vacuo* transfer following the MBE growth of the samples. As excitation, we used the 21.2 eV spectral line from a helium plasma lamp and the emitted photoemission electrons were detected by a Scienta Omicron DA 30L analyzer with energy resolution of 6 meV.

**STEM characterization and analysis**

Cross–section samples for STEM study were prepared on a FEI Helios Nanolab G4 dual-beam Focused Ion Beam (FIB) system with 30 keV Ga ions followed by ion-milling at 2 keV to removed damaged surface layers. Amorphous C and Pt were first deposited on the films to protect the surface from damage on exposure to the ion beam. STEM experiments were performed on an aberration–corrected FEI Titan G2 60-300 (S)TEM microscope, which is equipped with a CEOS

DCOR probe corrector, monochromator and a super-X energy dispersive X-ray (EDX) spectrometer. The microscope was operated at 200 and 300 keV with a probe current of 80 pA. HAADF-STEM images were acquired with the probe convergence angle of 25.5 mrad and the detector inner and outer collection angles of 55 and 200 mrad respectively. EDX elemental maps were acquired and analyzed using Bruker Esprit software. The lattice constants were obtained by using the Fourier transform atomic-resolution HAADF-STEM images.

**XRD and XPS characterization**

We carried out XRD measurements on X'Pert³ MRD operating in the reflection mode with Cu-Kα radiation (45kV, 40 mA) and diffracted beam monochromator, using a step scan mode with the step of 0.025°(2θ) and 0.88 s per step. The XPS experiments were performed using a Physical Electronics VersaProbe II instrument equipped with a monochromatic Al kα x-ray source (hν =1,486.6 eV) and a concentric hemispherical analyzer. Peaks were charge referenced to $CH_x$ band in the carbon 1s spectra at 284.8 eV. Measurements were made at a takeoff angle of 45° with respect to the sample surface plane. A model line-shape for the Te 3d spectrum was determined from an exfoliated, oxygen-free $Bi_2Te_3$ sample [53]. We assumed that the shapes would be very similar. The Cr 2p line-shape was modeled using reference Cr° spectra from the instrument vendor. Three sets of highly constrained doublets (1 each for Cr, Te° and $TeO_x$) were used on for the Cr 2p-Te 3d region. Quantification was done using instrumental relative sensitivity factors (RSFs) that account for the X-ray cross section and inelastic mean free path of the electrons. The analysis region was ~200 μm in diameter. The sapphire/$CrTe_2$ sample without capping layers as measured by XPS was transferred after removal from the MBE chamber into the XPS instrument within 5 minutes to minimize oxidation.

**Electrical transport characterization**

We performed electrical transport measurements in a Quantum Design physical properties measurement system (PPMS) in a Hall bar configuration. Hall bars with lateral dimensions of 1 mm x 0.5 mm were mechanically defined. The Hall resistance of the devices, $R_{yx}^*$, was measured as a function of magnetic field up to 3 T in the temperature range between 2 K and 300 K. To determine the anomalous Hall response at a given temperature, we first antisymmetrized the magnetic field dependence of the Hall resistance to remove the longitudinal resistance contribution; then, we subtracted the ordinary Hall resistance $R_{H_0}$: $R_{yx} = (R_{yx}^*(H,T) - R_{yx}^*(-H,T))/2 - R_{H_0}$. For samples that do not exhibit any THE, the AH resistance $R_{AHE}$ is then equal to $R_{yx}$. For samples that show a THE signal, the total transverse resistance can be written as $R_{yx} = R_{AHE} + R_{THE}$. The anomalous Hall conductivities $\sigma_{AH}$ are calculated as $\sigma_{AH} = \dfrac{\rho_{AHE}}{(\rho_{AHE}^2 + \rho_{xx}^2(H=0))}$.

**Pulsed switching measurement**

We carried out pulsed current-induced magnetization switching experiments in a PPMS using an external Keithley 2450 source meter and a Keithley 2182A nanovoltmeter. Before each switching attempt, the magnetization of the device was set and saturated in an initial state by applying a perpendicular magnetic field. In the pulse switching measurement, a train of pulses consisting of a 100 ms current pulse of varying magnitude followed by a 1500 ms pulse of 100 µA was applied under a magnetic field parallel to the current direction, during which we measured the anomalous Hall resistance of our system.

**Spin-torque ferromagnetic resonance measurement**

To further study the charge-to-spin conversion in the Dirac semimetal $ZrTe_2$ layer, we have performed ST-FMR in a $ZrTe_2$/permalloy heterostructure. Without breaking vacuum, we synthesized $ZrTe_2$ (5nm)/Py (4nm)/Al (4nm) heterostructures. These heterostructures were then patterned into 50 um x 10 um bars using standard lithography techniques including a two-step plasma etching process using $BCl_3$ and Ar as precursor gases. ST-FMR measurements were performed in these devices using a probe station equipped with a GMW 5201 projected field electromagnet, a Keysight E8257D analog signal generator and a Keithley 2182A nanovoltmeter. The spectrum was measured using a radiofrequency current ranging from 4 GHz to 6 GHz with an applied in plane magnetic field up to 1.6 KOe.

**DFT first principles calculation**

Spin-orbit-coupled (SOC) DFT calculations were implemented in the Vienna Ab-initio Simulation Package (VASP) [54–56] and Quantum Espresso [57]. The lattice constant for 1T-phase $CrTe_2$ and $ZrTe_2$ was obtained from the experimental result. The z-axis cell dimension is 15 Å for monolayer $CrTe_2$ to isolate a layer from its periodic images. The exchange-correlation is treated under GGA PBE approximation [58] with PAW method [59]. The energy cutoff in all calculations was 500 eV, and the k-point sampling was set as 16×16×1 and 16×16×10 centered at Γ for monolayer and bulk structures. The residual force after relaxation is smaller than 0.01 eV/A for all atoms. DFT+U method [60,61] is used in the calculation and the effective U is 2 eV to make results comparable to previous works [62,63]. Spin Hall conductivity is calculated based on kubo formula using the fitted Hamiltonian, as implemented in Wannier90 package [64].

**Acknowledgements**:

The MBE synthesis, ARPES, STM measurements and theoretical calculations were supported by the Penn State Two-Dimensional Crystal Consortium-Materials Innovation Platform (2DCC-MIP) under NSF Grant No. DMR-1539916 (YO, BZ, MS, TP, AR, VHC, NS). Transport measurements were carried out under support of the Institute for Quantum Matter under DOE EFRC grant DE-SC0019331 (RX, NS). The TEM, XRD, and SOT measurements were supported by SMART, one of seven centers of nCORE, a Semiconductor Research Corporation program, sponsored by the National Institute of Standards and Technology (NIST) (WY, YSH, SG, TL, KAM, NS). Parts of this work were carried out in the Characterization Facility, University of Minnesota, which receives partial support from the NSF through the MRSEC (Award Number DMR-2011401) and the NNCI (Award Number ECCS-2025124) programs (SG, KAM) We thank Jeffrey Shallenberger for assistance in XPS measurements and Hemian Yi for helpful discussions about ARPES measurements.

**Author Contributions**: YO and NS conceived the project and experiments. YO and MS grew the samples and perform the ARPES measurements. RX and MS performed the transport measurements. WY performed the pulsed current induced magnetization switching with the assistance from YO. MS, YH, TP and AR characterized the samples. SG conducted and analyzed the TEM data under the supervision of KAM. BZ performed the DFT band structure calculation under the supervision of VC. WJ and TL performed the spin Hall conductivity calculations. YO and NS wrote the manuscript with substantial contributions of all authors.

**Competing Interests:** The authors declare that they have no competing interests.

**Data availability**: All data needed to evaluate the conclusions in the paper are present in the paper and/or the Supplementary Materials. Additional data related to this paper may be requested from the authors.

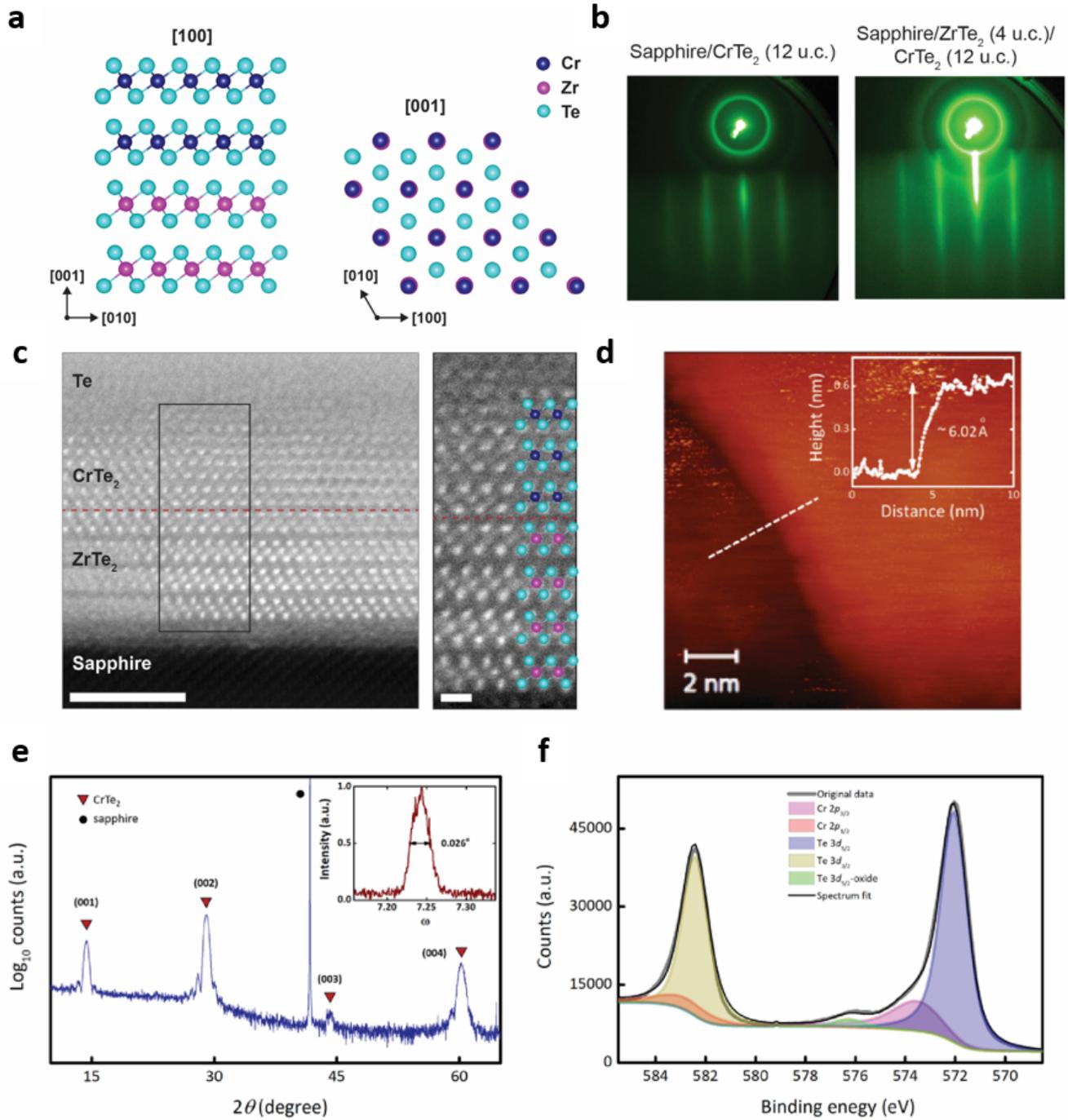

**Fig. 1. MBE and characterization of 1T-CrTe$_2$ thin films. a**, Schematics of the 1T phase of CrTe$_2$ grown on ZrTe$_2$. Note that [100], [010] and [001] correspond to [$\bar{1}2\bar{1}0$], [$2\bar{1}\bar{1}0$] and [0001]

in the 4 axis Miller-Bravais notation. **b,** RHEED patterns of the single layer of $CrTe_2$ (12 u.c.) and $ZrTe_2$ (4 u.c.)/$CrTe_2$ (12 u.c.) heterostructure. The substrate is sapphire in both cases and the electron beam is directed along the $[11\bar{2}0]$ orientation of sapphire. **c,** HAADF-STEM images of the $ZrTe_2$ (4 u.c.)/$CrTe_2$ (3 u.c.) heterostructure viewed in cross-section. The images have been low-pass filtered for clarity. Scale bar is 2 nm and 0.5 nm for the magnified panel. **d,** STM image of a $ZrTe_2$ (4 u.c.)/$CrTe_2$ (1 u.c.) sample. The line scan shows the thickness of the 1 u.c. $CrTe_2$. **e,** XRD $2\theta$ scan of a 12 u.c. $CrTe_2$ thin film. **f,** XPS spectrum of an 18 u.c. thick $CrTe_2$ film. The small Te oxide shoulder is due to absence of a capping layer in this sample.

**Figure 2**

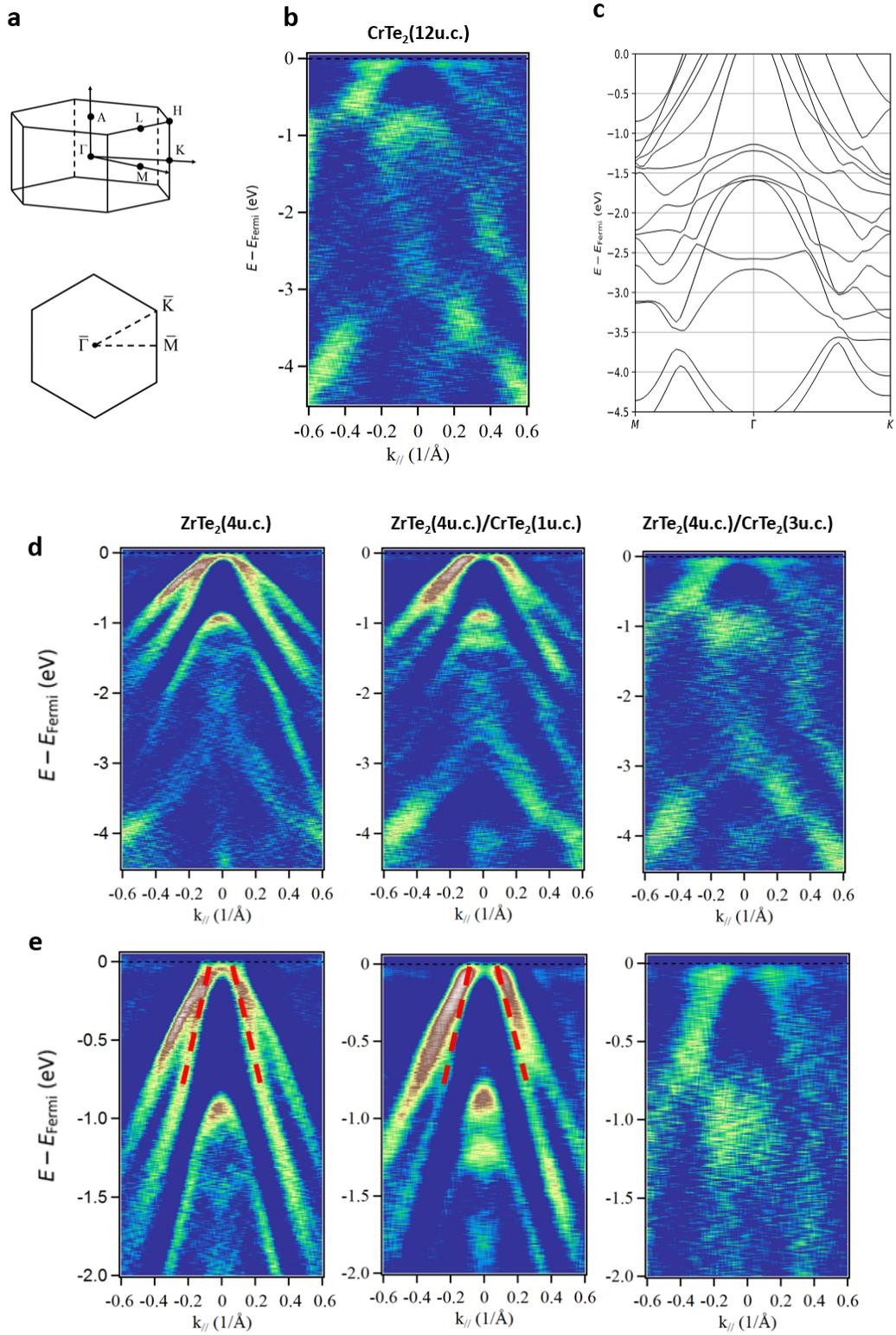

**Fig. 2. ARPES measurements and DFT calculation of the band structure of 1T-CrTe$_2$ thin films. a,** Schematic of the bulk and projected Brillouin zones of CrTe$_2$. **b,** ARPES spectrum of a single layer of 12 u.c. CrTe$_2$ in the $\bar{\Gamma}-\bar{M}$ direction. **c,** DFT calculation of the band structure of bulk CrTe$_2$. The M and K points are 0.92 and 1.06 Å$^{-1}$ respectively. **d** and **e**, ARPES spectrum of the 4 u.c. ZrTe$_2$ (left), ZrTe$_2$ (4 u.c.)/CrTe$_2$ (1 u.c.) (middle) and ZrTe$_2$ (4 u.c.)/CrTe$_2$ (3 u.c.) (right) with a larger (d) and smaller (e) binding energy scale. All the ARPES data are taken at 300 K with 21.2 eV excitation from a He lamp. To more clearly highlight the measured band dispersion, we present all plots as second-derivatives with respect to the energy. The red dashed lines are guides to the eyes for the Dirac dispersion in ZrTe$_2$.

**Figure 3**

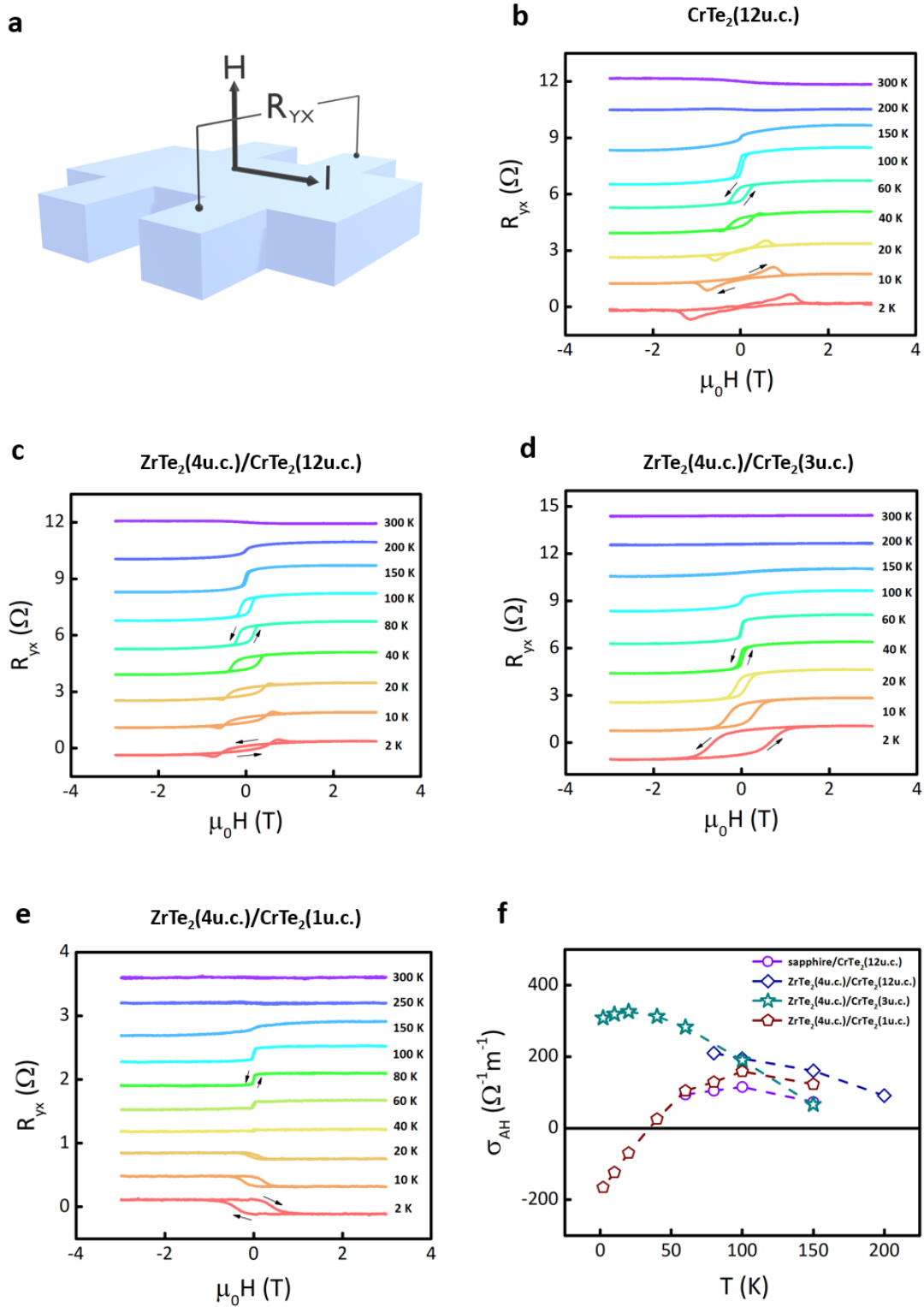

**Fig. 3. Anomalous Hall resistance measurements on CrTe$_2$ layers and ZrTe$_2$/CrTe$_2$ heterostructure. a**, Schematics of the Hall bar. **b** to **e,** Anomalous Hall resistance (AHR) of the 12 u.c. CrTe$_2$ (b), ZrTe$_2$ (4 u.c.)/CrTe$_2$ (12 u.c.) (c), ZrTe$_2$ (4 u.c.)/CrTe$_2$ (3 u.c.) (d) and ZrTe$_2$ (4 u.c.)/CrTe$_2$ (1 u.c.) (e). The AH resistance data has been offset for clarity. The arrows denote the field sweep directions. **f,** Anomalous Hall conductivity of the samples shown from (b) to (e).

**Figure 4**

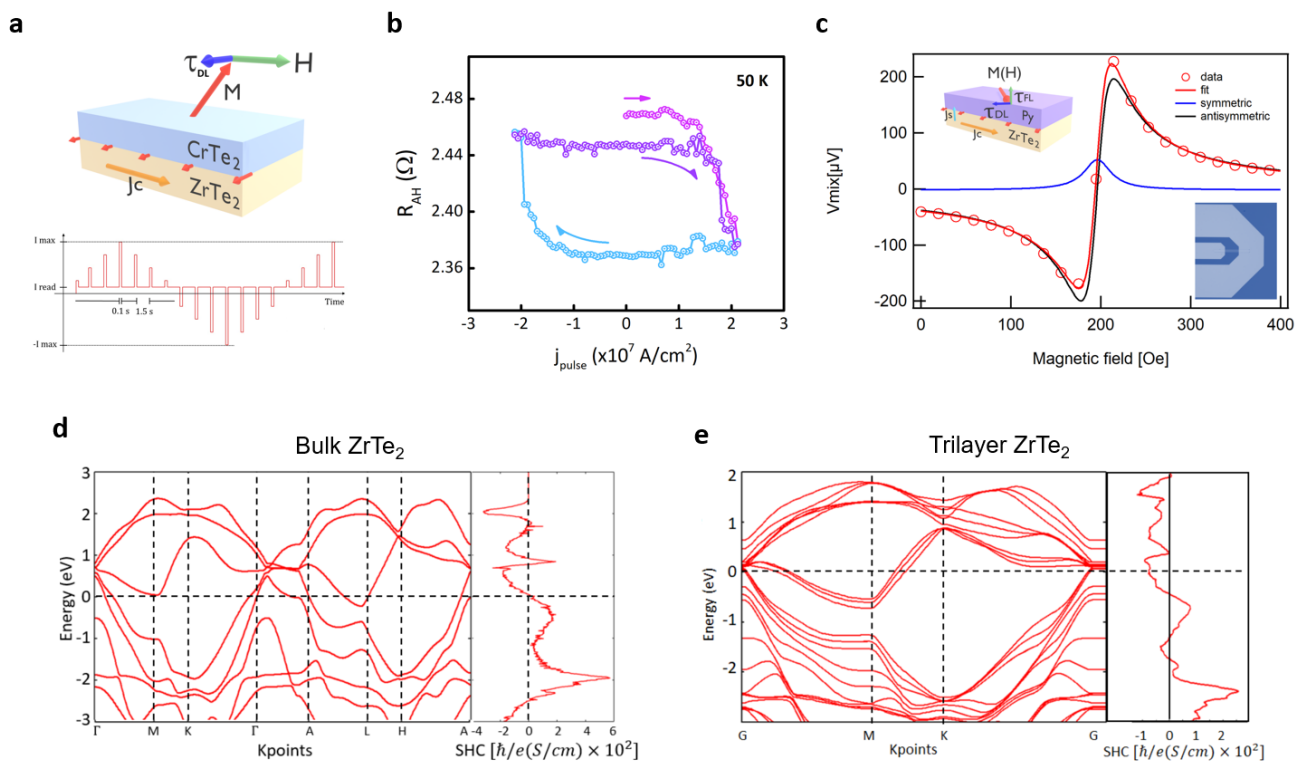

**Fig. 4. Pulsed current-induced magnetization switching of an ultrathin ZrTe₂/CrTe₂ heterostructure device and SOT characterizations of ZrTe₂. a,** Schematics of the SOT assisted magnetization switching in the ZrTe₂/CrTe₂ heterostructure and the writing and reading pulse current sequence. **b,** Pulse current induced magnetization switching of a ZrTe₂ (8u.c.)/CrTe₂ (3u.c.) at 50 K under an external magnetic field (700 Oe) applied in plane parallel to the current direction. **c**, ST-FMR spectra of a ZrTe₂/Py bilayer heterostructure. Inset: Optical microscope image and schematic of the ST-FMR device. **d** and **e**, DFT calculated band structure and spin Hall conductivity of ZrTe₂ in the bulk and thin film form, respectively.

Supplemental Information

ZrTe2/CrTe2: an epitaxial van der Waals platform for spintronics

Yongxi Ou, Wilson Yanez, Run Xiao, Max Stanley, Supriya Ghosh, Boyang Zheng, Wei Jiang, Yu-Sheng Huang, Timothy Pillsbury, Anthony Richardella, Chaoxing Liu, Tony Low, Vincent H. Crespi, K. Andre Mkhoyan, Nitin Samarth

**S1 Characterization of the MBE-grown topological semimetal ZrTe$_2$**

MBE-grown pristine ZrTe$_2$ has been reported to exhibit massless Dirac fermions in its band dispersion with a Dirac cone at the gamma point, making it an interesting candidate for VdW Dirac semimetals [37,38]. We have grown ZrTe$_2$ on sapphire substrates by following the process described in the Methods sections. Here we provide more details about characterization of our MBE-grown ZrTe$_2$ films. Figure S1A shows the XRD spectrum of a sapphire/ZrTe$_2$ (6u.c.) sample indicating the peaks of ZrTe$_2$ along the out-of-plane (001) growth direction, in good agreement with the literature [38]. Figure S1B (same as Fig. 2 in the main text) shows ARPES measurements of a 4 u.c. thick ZrTe$_2$ film. The parabolic valence band and the linear Dirac dispersion are visible in Fig. S1B. We note that the chemical potential lies below the Dirac point in this sample. This is in good agreement with the results in Ref.[38]. However, this makes direct visualization of the Dirac point difficult. A shift in the Fermi level closer to the Dirac point may be achieved by growing the sample on another type of substrate (e.g. on InAs)[37] and/or via post-growth annealing treatment makes the sample n-doped due to the introduction of Te vacancies. We adopted both methods by annealing a ZrTe$_2$ sample grown on an epi-graphene substrate at 600 °C for 2 hours. As shown in Fig. S1C, the sample indeed became more n-type doped after the annealing treatment resulting in a clearer Dirac point indicated by the converged linear dispersion. More details about the ZrTe$_2$ synthesis and characterization will be provided in future work. As a control measurement, we also measured the Hall effect in a sapphire/ZrTe$_2$ (4u.c.) sample at various temperatures (Fig. S1D). The lack of any AHE signal in ZrTe$_2$ confirms that the ferromagnetic response in the ZrTe$_2$/CrTe$_2$ samples described in the main text originates in the 1T-CrTe$_2$ layer.

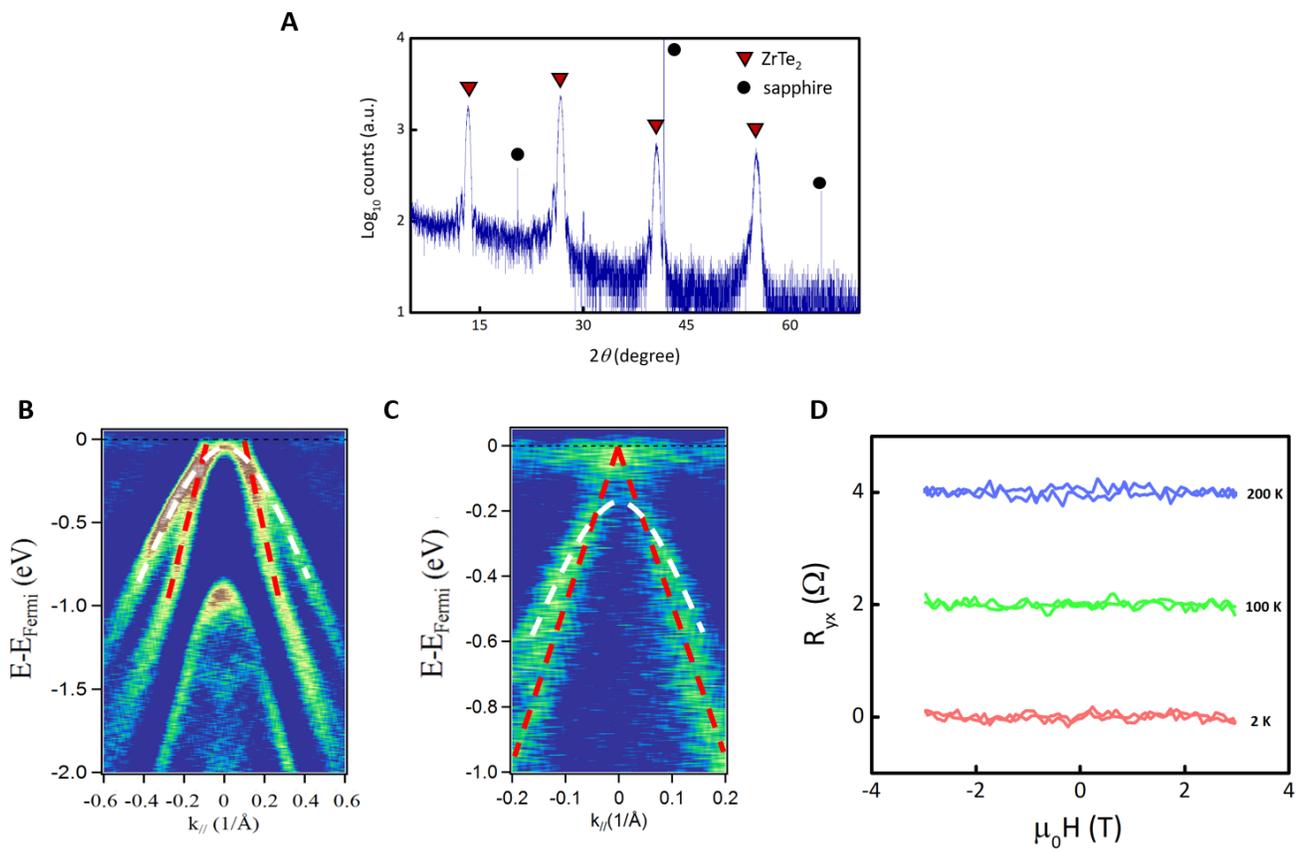

**Figure S1**. (A) XRD $2\theta$ scan of a sapphire/ZrTe$_2$(6u.c.) sample. (B and C) ARPES spectrum on (B) a sapphire/ZrTe$_2$(4u.c.) sample and (C) an epi-graphene/ZrTe$_2$(10u.c.) sample at room temperature (light source: 21.2 eV). Both ARPES plots are second-derivatives. The red and white

dashed lines are guidance to the eyes. (D) Anomalous Hall resistance of a sapphire/ZrTe$_2$(4u.c.) thin film at various temperature. The AH resistance data has been offset for clarity.

**S2 STEM imaging and energy dispersive X-ray (EDX) spectroscopy of the 1T-CrTe$_2$**

Figure S2A shows the HAADF-STEM image of a 1T-CrTe$_2$ thin film. The elemental distribution and composition in these MBE-grown CrTe$_2$ thin films were evaluated via the STEM-EDX mapping (Fig. S2B). The EDX line-scan in Fig. 2C shows that Cr atoms are clearly confined to the CrTe$_2$ layer and no oxygen signal is seen in the material layer, indicating it is pristine. From this EDX mapping analysis, we determined the Cr to Te ratio in the grown CrTe$_2$ thin film to be 0.53, which is very close to 1:2, indicating little (if any) Cr intercalation into the films (Fig. S2C).

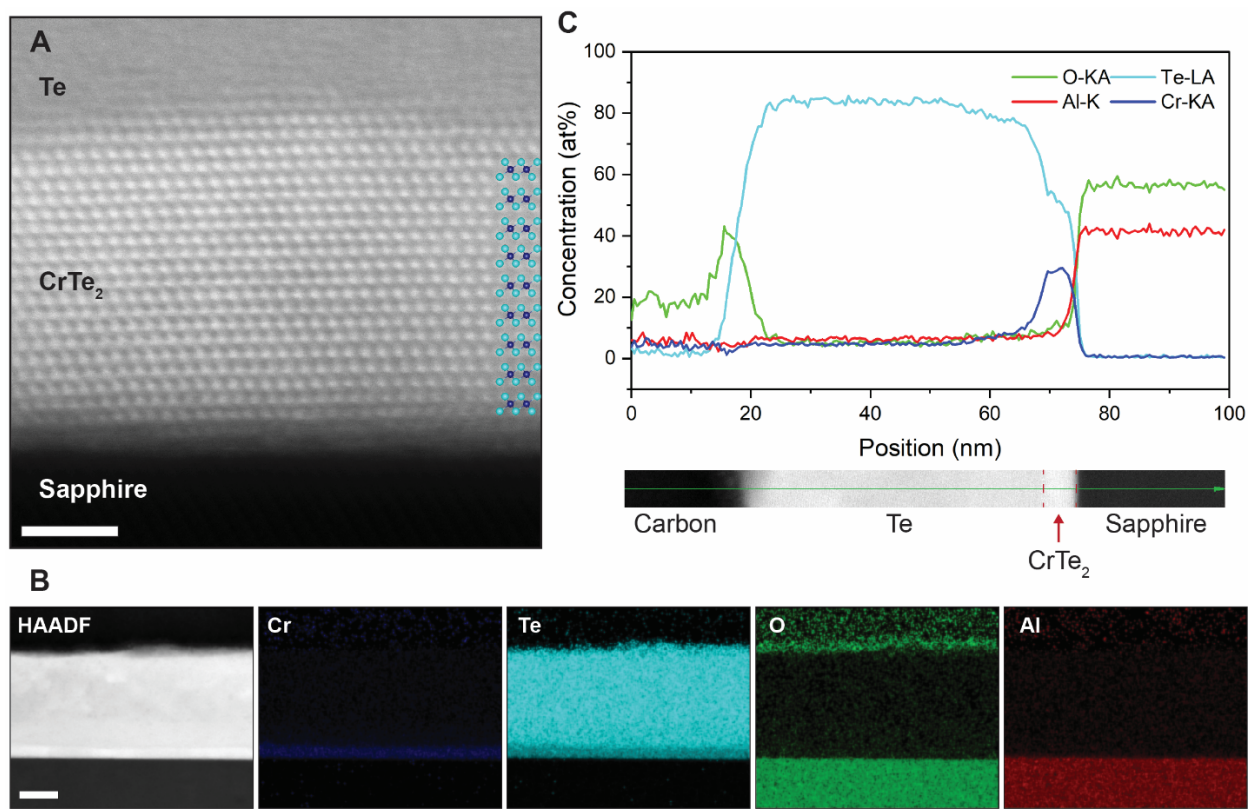

**Figure S2**. (A) HAADF-STEM images of the sapphire/CrTe$_2$(12u.c.) sample. The image has been low-pass filtered for clarity. Scale bar is 2 nm. (B) A set of STEM-EDX elemental maps of a

sapphire/CrTe$_2$(12u.c.)/Te sample shown along with the HAADF-STEM image acquired with the maps. Scale bar is 20 nm. (C) A line-scan of the EDX signal obtained from the region in the panel (B) showing the relative composition of the elements present in the film.

## S3 XRD reciprocal mapping

A reciprocal space map (RSM) was taken on the X'Pert³ MRD around the symmetric $CrTe_2$ (002) peak using a PIXcel 3D detector in static line mode (shown in Fig. S3). The resulting RSM shows little mosaic disorder, consistent with the measured rocking curve width. Attempts to measure an RSM on an asymmetric peak were unsuccessful because the film was so thin; thus, the amount of strain could not be determined.

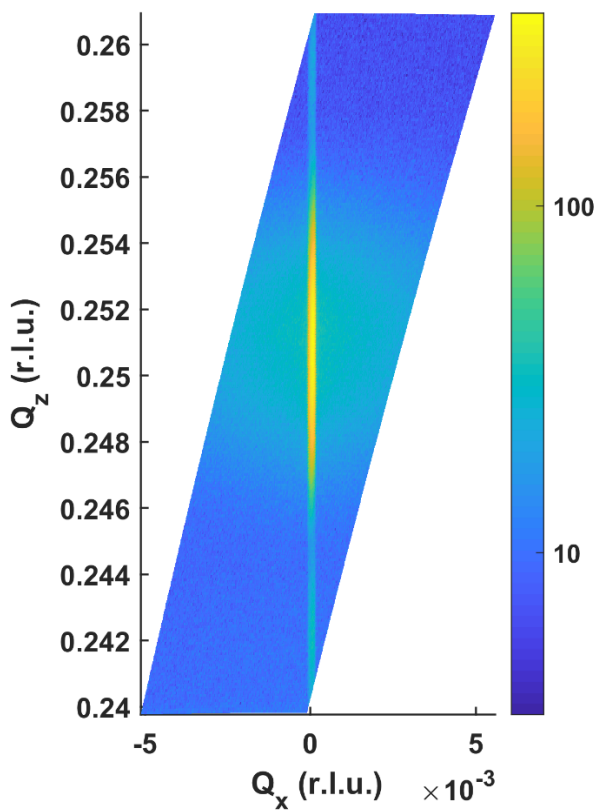

**Figure S3**. Reciprocal space map of the $CrTe_2$ (002) peak of a sapphire/$CrTe_2$(12u.c.)/Te sample.

**S4 Comparison between the ARPES results and the DFT calculation**

The comparison of the ARPES spectrum for the 12u.c. $CrTe_2$ thin film makes use of the computed Fermi level from PBE-based density functional theory including spin-orbit coupling and finite Hubbard U, using the computed band structure of the bulk plotted along the Γ-M-K plane. Here we consider effects related to $k_z$ dispersion and also discuss the possibility of charge transfer or extrinsic doping shifting the Fermi level. Figure S4A and S4B shows the band structure of bulk 1T-$CrTe_2$ calculated by DFT, with Figure S4B providing Brillouin zone sampling that also extends in the *z* direction (Γ to A). This $k_z$ dispersion should lift slightly the band at Γ around –1.2 eV and also drop slightly the hole-like bands around the Fermi level; both these changes would correct small discrepancies in the comparison of theory and experiment. While bulk calculations can give a sense of the effects of *z*-axis dispersion, an explicit multilayer calculation for multilayer $CrTe_2$ can provide a more direct sense of these effects. Figures S4C, D and E show the bands of a 3 u.c. $CrTe_2$ slab, color-coded so that bands that project onto the top and bottom layers are red and blue (combining to purple) while those projecting onto the middle layer are green. We focus on the purple (i.e. near-surface) bands, and show three possible choices of Fermi level corresponding to shifts of 0, 0.75 and 1 eV. Both non-zero shifts place one or more purple bands into the signal-free region around half a volt below the experimental Fermi level; these discrepancies suggest that the unshifted Fermi level presented in the main text is the most likely band alignment.

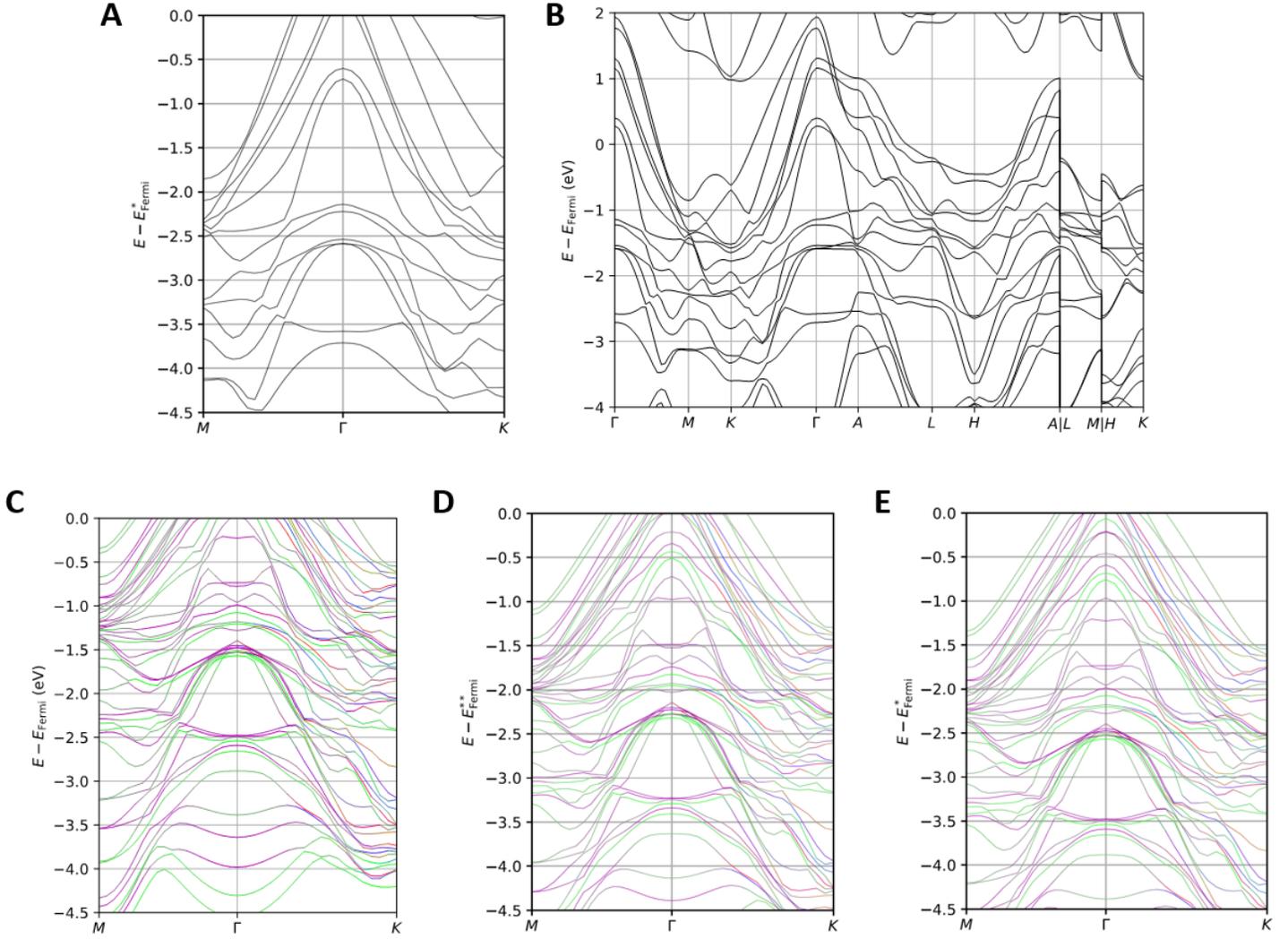

**Figure S4**. (A and B) DFT calculated band structures for bulk $CrTe_2$. (C-E) DFT results for 3 u.c. of $CrTe_2$ with different energy windows. The color coding comes from the projection to different layers. Red for the 1st layer, blue for the 2nd and green for the 3rd. Purple is the combination of red and blue. The M and K points are 0.92 and 1.06 Å$^{-1}$ respectively. $E^*_{Fermi} = E_{Fermi} + 1$, $E^{**}_{Fermi} = E_{Fermi} + 0.75$.

## S5 DFT calculations of the magnetic anisotropy energy

To understand the out-of-plane magnetic anisotropy in our $CrTe_2$ thin films, we calculated the magnetic anisotropy energy (MAE) for $CrTe_2$ as a function of the in-plane lattice constant (*a*) with different electron doping levels in Fig.S5. The MAE is demonstrated in the form of $E_{in}$-$E_{out}$ where $E_{in}$ could either be the energy for spins lying in the x or y-directions. The positive value in Fig. S5 indicates an out-of-plane easy axis. Our calculated results indicate that the energy difference between out-of-plane and in-plane directions is much larger than that between the two in-plane directions, *x* and *y*, in good agreement with previous works [62,63]. While the observed out-of-plane anisotropy in our $CrTe_2$ samples is consistent with existing experimental works [32,34], we notice that MAE can be quite sensitive to the lattice constant *a*, electron doping level and the choice of Hubbard U [34], which suggests that the easy axis direction is hard to be determined from the theoretical point of view due to its sensitivity to the experimental parameters.

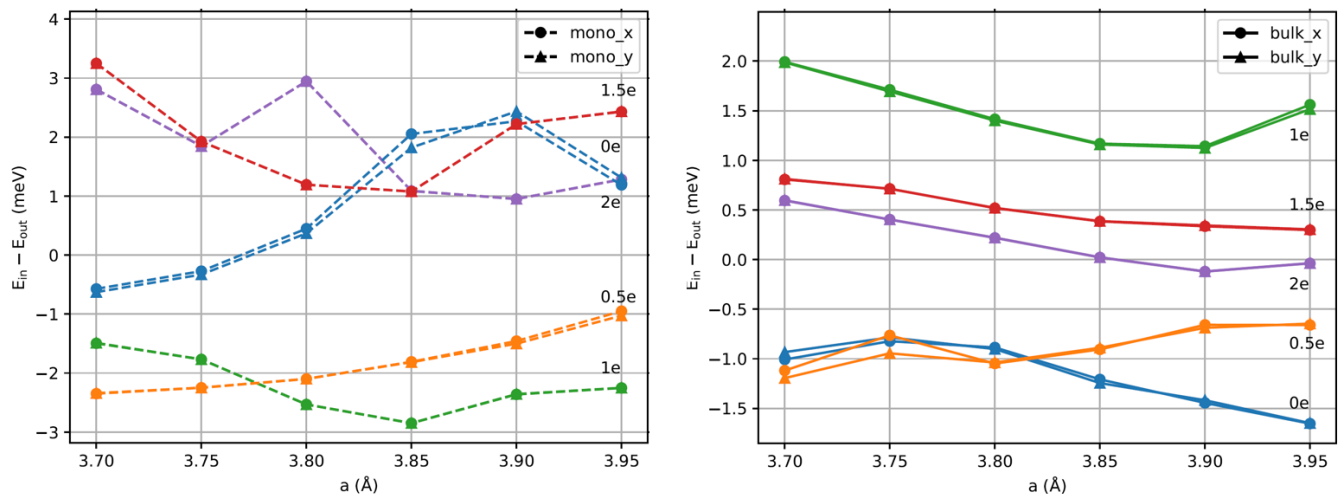

**Figure S5**. MAE as a function of lattice constant *a* with different electron doping levels. $E_{out}$ is the energy for spins along out-of-plane (z) direction, and $E_{in}$ is for the in-plane (x/y) direction.

"mono"/"bulk" refers to monolayer/bulk calculations. The lattice constant $c$ is fixed at 15 Å for monolayer and 6.02 Å for bulk calculations.

## S6 Longitudinal resistivity measurements of the 1T-CrTe₂ epilayer

Figure S6 shows the temperature dependent longitudinal resistivity of a sapphire/CrTe$_2$(12u.c.)/Te sample measured in a PPMS apparatus in a four-terminal configuration on a 1 mm x 0.5 mm Hall bar. At room temperature (300K), the resistivity of the CrTe$_2$ is $\rho_{xx} \approx 594\ \mu\Omega$.cm.

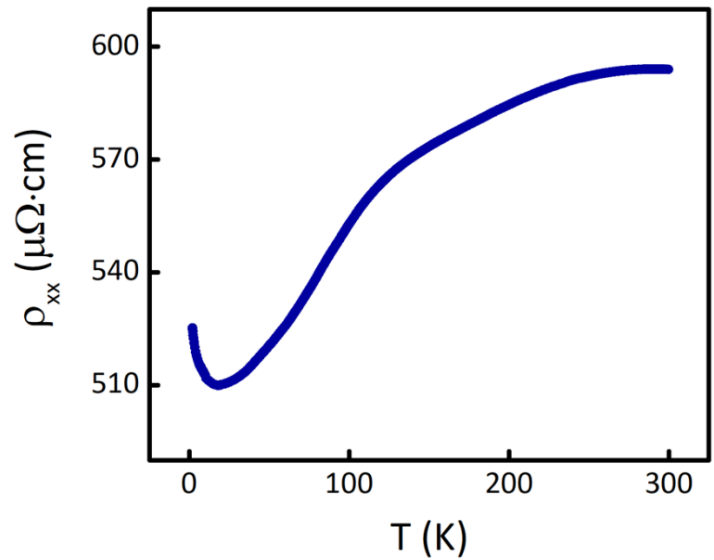

**Figure S6**. Temperature dependent longitudinal resistivity of a sapphire/CrTe$_2$(12u.c.)/Te sample.

## S7 Attempting fits to the THE signal by using the two AHE model and field-angle dependence measurements

As we showed in Fig. 3 in the main text, the AHE measured in both 1T-CrTe$_2$ layer and ZrTe$_2$/CrTe$_2$ heterostructures can show behavior consistent with a THE arising from Berry curvature in real space. However, such behavior can also arise from coexisting anomalous Hall signals with opposite signs. Such an interpretation has been used to explain an unconventional THE-like anomalous Hall resistance in magnetically-doped topological insulator thin films and in oxide interfaces [65,66]. To test this scenario, we carried out a similar fitting process to our THE signal by assuming two AH effects following Ref.[45,65]. For this purpose, the signals were fitted to:

$$R_{yx}(H) = R_{AH1} \tanh\left(\frac{H \pm H_{c1}}{H_{01}}\right) + R_{AH2} \tanh\left(\frac{H \pm H_{c2}}{H_{02}}\right).$$

Here, $R_{AH1}$ ($R_{AH2}$) is the first (second) anomalous Hall resistance contribution with $H_{c1}$ ($H_{c2}$) being the coercivity and $H_{01}$ ($H_{02}$) being a scaling factor, respectively. Examples of the best fits compared to the original data are plotted in Fig. S7A, where we see that the fits do not reproduce all the details of the original data. Fig. S7B summarizes the fitted parameters as a function of temperature up to where the THE signal disappeared in our transport measurement. The variation of the fitting parameters with temperature does not yield a physically meaningful picture: for example, while $R_{AH2}$ increases with decreasing temperature, $R_{AH1}$ however *decreases* with decreasing temperature even though its corresponding coercive field $H_{c1}$ and scaling factor $H_{01}$ do not decrease. The scaling factor $H_{02}$ also shows as a non-monotonic behavior. This suggests that the competing AHE picture is probably not a valid one.

We also performed a field-angle dependent measurement to determine the influence of the magnetic field's direction on the THE signal. Figure S7C shows that the THE signal is relatively insensitive to the field angle up to around 45° and then disappears at higher field angles. The field-angle behavior of the THE effect is similar to what is observed in Ref.[45] and it may be related to the energy rebalancing process as the magnetic field is rotated. As we mention in the main text, direct magnetic imaging such as low temperature MFM and Lorentz TEM measurements will be needed to further determine the nature of the THE signal observed in our 1T-CrTe$_2$ thin films.

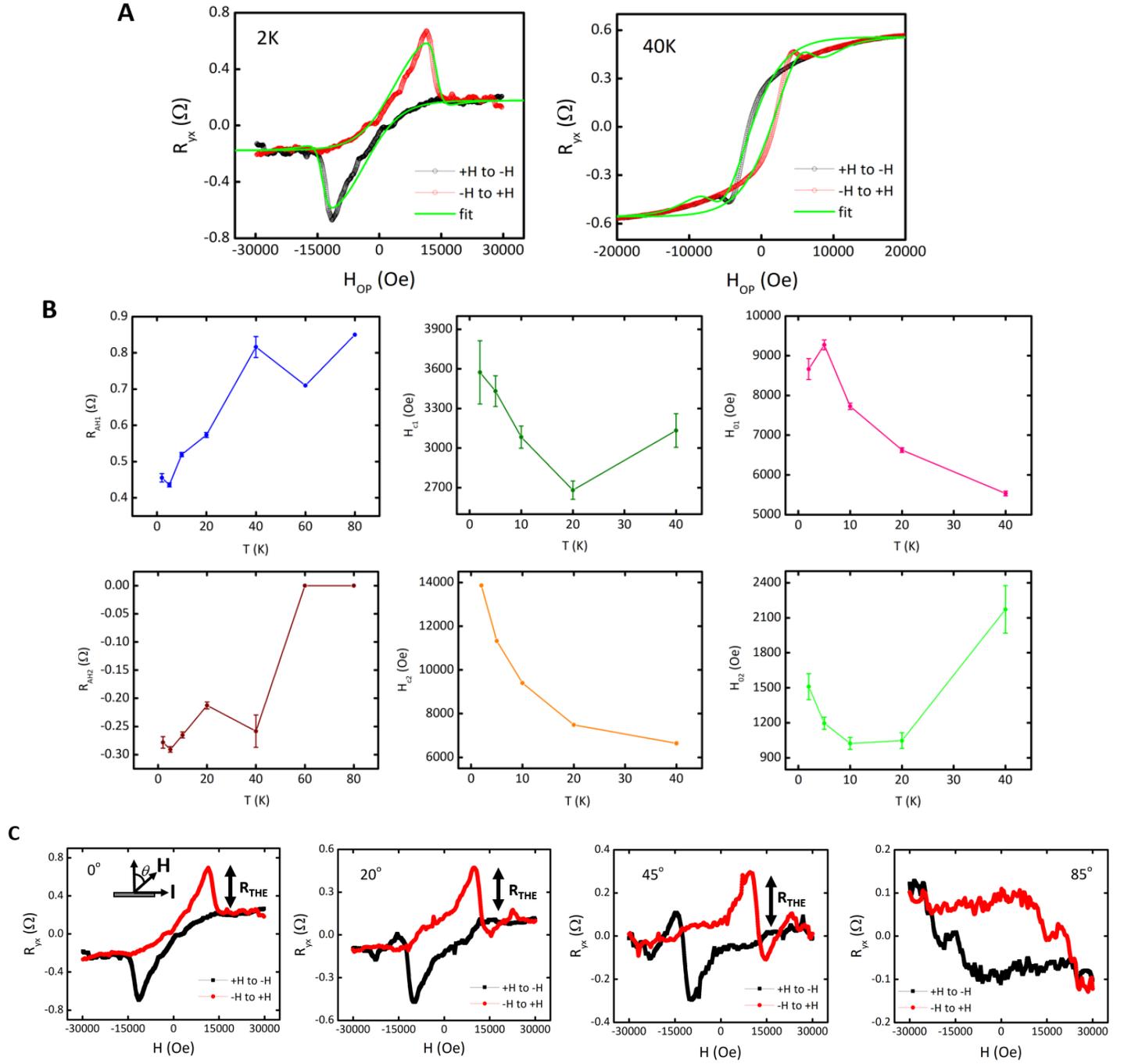

**Figure S7**. (A) Fits to the Hall signals of the sapphire/CrTe$_2$(12u.c.) sample using the bi-AHE models. (B) Temperature dependent fitting parameters attained from the bi-AHE fits. (C) Field-angle dependence of the Hall signals of the sapphire/CrTe$_2$(12u.c.) sample at 2K.

## S8. Current-induced magnetization switching in ZrTe$_2$/CrTe$_2$

By measuring the anomalous Hall resistance to determine the magnetization state of CrTe$_2$, we tested the current-induced magnetization switching of the ZrTe$_2$(8u.c.)/CrTe$_2$(3u.c.) devices between 10K and 90K under various external in-plane magnetic field along the electrical current direction, as shown in Fig. S8. The step-like switching edges during most of these switching attempts are consistent with the domain nucleation and domain wall motion, probably due to the micro-scale size of the devices.

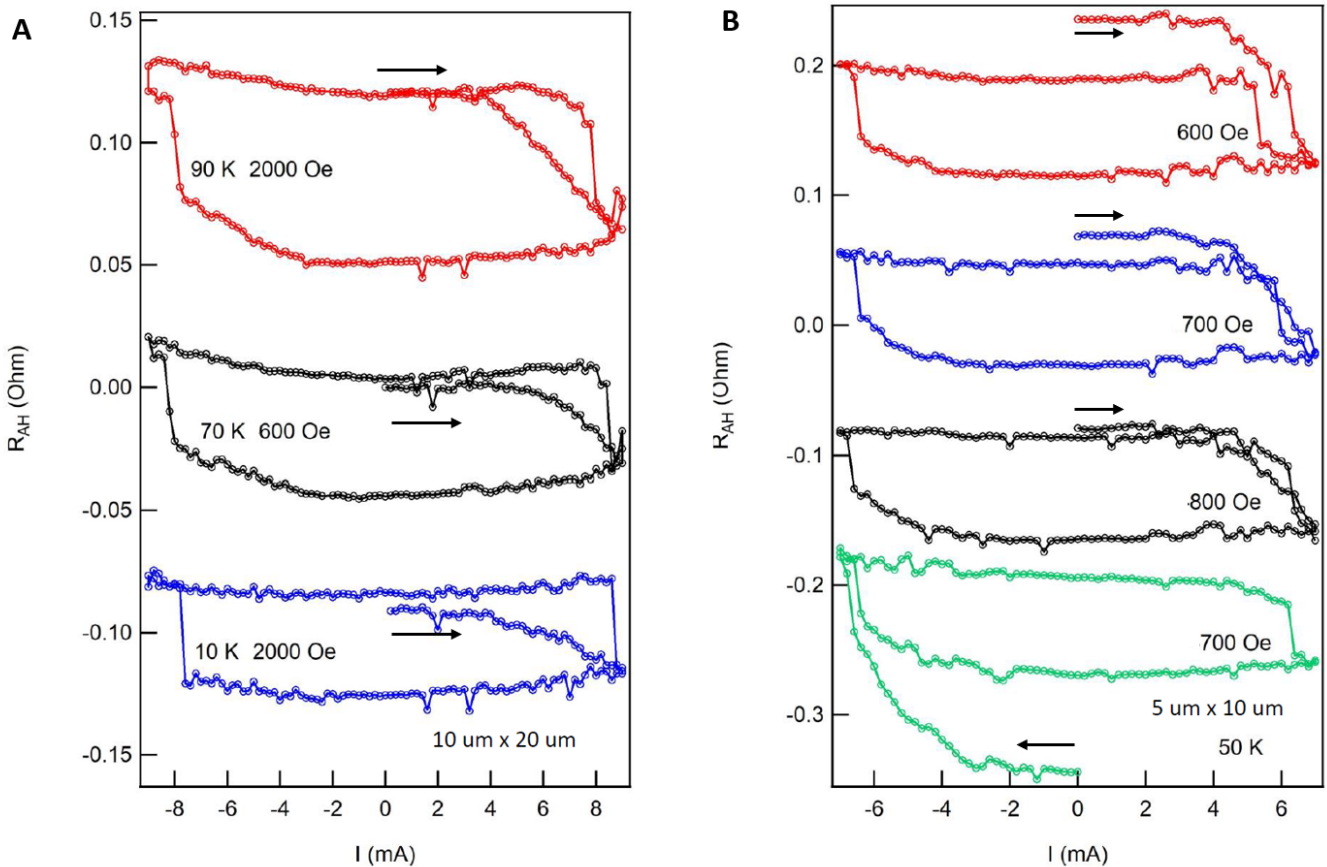

**Figure S8**. (A) Current-induced magnetization switching measurements at 10K, 70K and 90K. (B) Current-induced magnetization switching measurements at 50K under various in-plane magnetic field. The arrows indicate the starting positions.

## S9 DFT calculations of SHC of ZrTe$_2$ thin films

Figure S9 shows the atomic structure, band structure, and energy dependent SHC of monolayer and bilayer ZrTe$_2$ thin films. The band structure of monolayer and bilayer systems is similar to that of the trilayer system. The Fermi level in thin films is clearly shifted to higher energy compared to the bulk phase, which is caused by the change of band dispersion especially along the out-of-plane direction due to the quantum confinement effect.

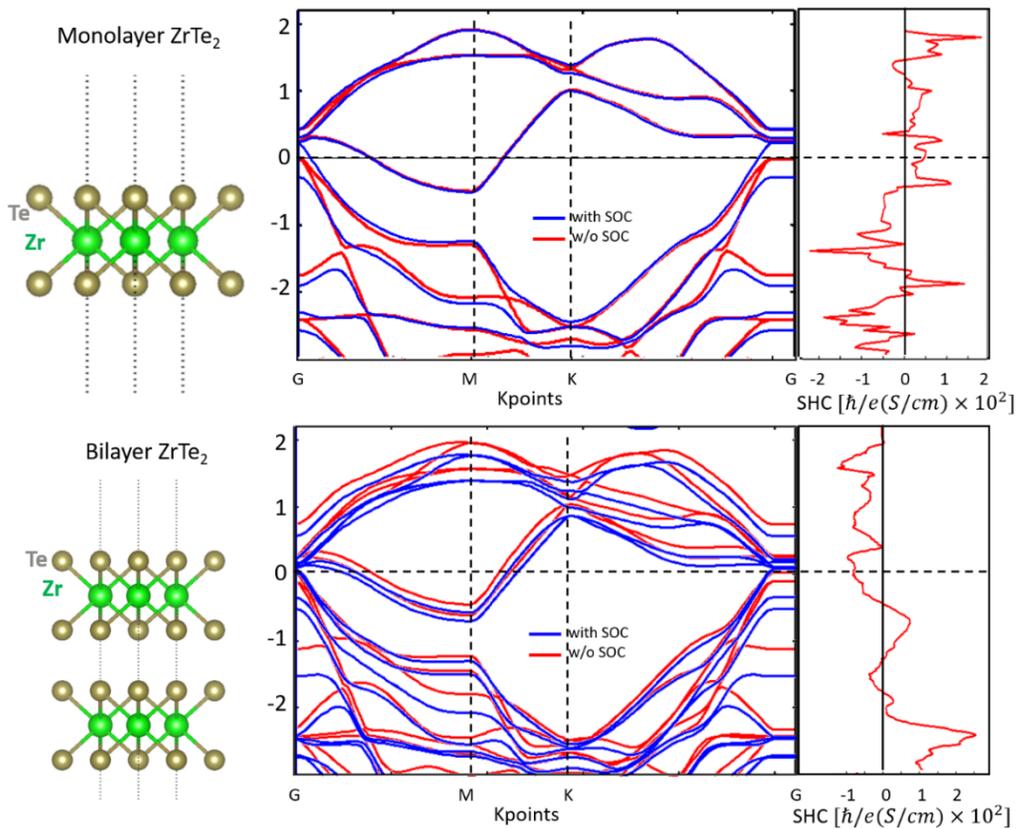

**Figure S9**. DFT calculated band structure and spin Hall conductivity of the monolayer (upper panel) and bilayer (lower panel) ZrTe$_2$ thin films, respectively. Blue and red lines corresponding to band structure with and without spin-orbit coupling, respectively.